# The CMS Event Builder

V. Brigljevic, G. Bruno, E. Cano, S. Cittolin, A. Csilling, D. Gigi, F. Glege, R. Gomez-Reino,
M. Gulmini[1], J. Gutleber, C. Jacobs, M. Kozlovszky, H. Larsen, I. Magrans de Abril, F. Meijers[*],
E. Meschi, S. Murray, A. Oh, L. Orsini, L. Pollet, A. Racz, D. Samyn, P. Scharff-Hansen,
C. Schwick, P. Sphicas[2]
*CERN, European Organization for Nuclear Research, Geneva, Switzerland*
[1]*Also at Laboratori Nazionali di Legnaro dell'INFN, Legnaro, Italy*
[2]*Also at MIT, Cambridge, USA and University of Athens, Greece*

V. O' Dell, I. Suzuki
*Fermi National Accelerator Laboratory, Batavia, Illinois, USA*

L. Berti, G. Maron, N. Toniolo, L. Zangrando
*Laboratori Nazionali di Legnaro dell'INFN, Legnaro, Italy*

A. Ninane
*Université Catholique de Louvain, Louvain-la-Neuve, Belgium*

S. Erhan
*University of California, Los Angeles, California, USA*

S. Bhattacharya, J. Branson
*University of California, San Diego, California, USA*

[*] Plenary talk presented by Frans Meijers (frans.meijers@cern.ch)

The data acquisition system of the CMS experiment at the Large Hadron Collider will employ an event builder which will combine data from about 500 data sources into full events at an aggregate throughput of 100 GByte/s. Several architectures and switch technologies have been evaluated for the DAQ Technical Design Report by measurements with test benches and by simulation. This paper describes studies of an EVB test-bench based on 64 PCs acting as data sources and data consumers and employing both Gigabit Ethernet and Myrinet technologies as the interconnect. In the case of Ethernet, protocols based on Layer-2 frames and on TCP/IP are evaluated. Results from ongoing studies, including measurements on throughput and scaling are presented. The architecture of the baseline CMS event builder will be outlined. The event builder is organised into two stages with intelligent buffers in between. The first stage contains 64 switches performing a first level of data concentration by building super-fragments from fragments of 8 data sources. The second stage combines the 64 super-fragments into full events. This architecture allows installation of the second stage of the event builder in steps, with the overall throughput scaling linearly with the number of switches in the second stage. Possible implementations of the components of the event builder are discussed and the expected performance of the full event builder is outlined.

## 1. INTRODUCTION

CMS is one of the experiments currently under construction for the future Large Hadron Collider (LHC) at CERN and is foreseen to start operation in 2007. The CMS experiment [1] will employ a general-purpose detector comprising electromagnetic and hadronic calorimeters, a muon system and tracking detectors. The main parameters of the Trigger and Data Acquisition (TriDAS) system in the case of proton-proton collisions at the design LHC luminosity of $10^{34}\,cm^{-2}s^{-1}$ are summarised in Table 1.

In order to reduce the event rate from the 40 MHz LHC beam crossing to an acceptable rate of O(100) Hz for archiving and later offline analysis, a rejection power of $O(10^5)$ is required online. The online event selection is done in two steps: the Level-1 Trigger implemented in hardware and the High Level Trigger (HLT) implemented in software.

Table 1: Nominal parameters of the CMS DAQ system.

| Parameter | Value |
|---|---|
| Beam crossing rate | 40 MHz |
| Level-1 Trigger rate | 100 kHz |
| Number of front-ends | 700 |
| Event Size | 1 MByte |
| Event Builder throughput | 100 GByte/s |
| Maximum rate after HLT | O(100) Hz |
| HLT computing power | $10^6$ SpecInt95 |
| Data production | 10 TByte/day |

The architecture of the Data Acquisition (DAQ) system is schematically shown in Figure 1. The Level-1 Trigger [2] operates at the LHC beam crossing rate using information from the calorimeters and muon system and is expected to reduce the event rate to 100 kHz. During the Level-1 Trigger latency of ~3 μs, the data from all detector channels are stored in front-end pipeline memories. For the events accepted by the Level-1 Trigger, all data are read out in parallel into the Front-End Drivers (FEDs). The estimated data volume after data reduction in the front-ends [3] is given in Table 2. There are roughly 700 FED modules, each carrying 1 to 2 kByte (kB) of data on average per event. Considering the uncertainty in the size estimate, a total average event size of 1 MB is used as a working assumption.

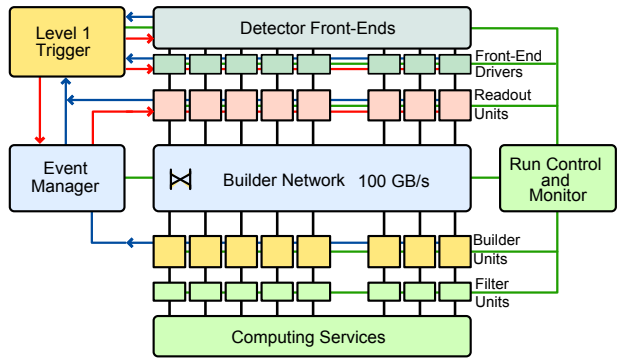

Figure 1: The CMS data acquisition architecture. The Event Builder comprises the Readout Units, the Builder Units, the Event Manager and the Builder Network connecting them.

The Event Builder (EVB) is a central component of the DAQ system. An event enters the system as a set of fragments distributed over the FEDs. It is the task of the EVB to collect the fragments of an event, assemble them and send the full event to a single processing unit. To this end, a builder network connects ~500 Readout Units (RUs) to ~500 Builder Units (BUs). The RUs are read out and buffer event fragments from the FEDs. The buffer capacity in the RUs corresponds to roughly 1 s. A BU collects the event fragments belonging to the same event from all RUs and buffers the full event. Each BU assembles a different event in parallel. With an average event size of 1 MB, the EVB network requires an effective aggregate throughput of 100 GB/s. The event flow through the EVB is supervised by an event manager (EVM), which communicates with the Level-1 Trigger and controls the event building process by mediating control messages between RUs and BUs.

Table 2: Detector readout characteristics.

| Detector | Nb of Channels | Nb of FEDs | Data Size (kByte) |
|---|---|---|---|
| Tracker Pixel | ~44M | 38 | 72 |
| Tracker Strips | ~9M | 440 | 300 |
| Preshower | 144384 | 47 | 110 |
| ECAL | 82728 | 54 | 100 |
| HCAL | 9072 | 32 | 64 |
| Muons CSC | ~500k | 8 | 12 |
| Muons RPC | 192k | 6 | 2 |
| Muons DT | 195k | 5 | 8 |
| Trigger | | 6 | 11 |
| **Total** | | **636** | **679** |

For simplicity, the EVB has been described above as a monolithic switch network connecting ~500 data sources (the RUs) to ~500 data consumers (the BUs). However, a number of implementations are possible, including a two-stage event builder.

Given the event size of 1 MB, the event rate of 100 kHz and the number of RU and BU nodes of 500, the ports of all RU and BU nodes must operate at an effective speed of about 2 Gbit/s. If the switch network does not support this effective speed, it is still possible to achieve the 2 Gbit/s using multiple parallel switch networks, called "rails". In this case, each node accommodates one port per rail.

As the EVB is based on a switch network, a 100% utilisation of the network can only be achieved if all inputs are balanced. Thus, it is important to have similar fragment sizes on average across all switch inputs.

The High-Level Trigger system consists of a series of reconstruction and selection algorithms from the offline environment. They are designed to reduce the maximum event rate of 100 kHz to O(100)Hz events forwarded to storage. The HLT operates on events fully assembled by the EVB and hence can use all detector data with full granularity and resolution. It is only limited by the available processing power and the quality of the online calibration. The HLT algorithm will run on a large farm of commodity processing nodes called Filter Units (FUs). The FUs are served with events by the BUs, where a BU is typically connected to a few FUs via a dedicated network. In addition to the data reduction, the HLT farm will provide quasi-real-time feedback on the detector performance and physics data quality. The CPU power required for the HLT farm has been estimated running algorithms providing the required rate reduction on a large sample of simulated events. Based on current processing times[1], a total computing power for the HLT farm of roughly $10^6$ SpecInt95 for O(1000) CPUs in the year 2007 is needed.

All components of the DAQ are supervised by the Run Control and Monitor System (RCMS) [3,4] of the experiment.

The design of the overall CMS Data Acquisition and High Level Trigger has been described in detail in the Technical Design Report [3]. This paper summarises the design of the Event Builder. Section 2 reports on event builder studies with test benches. Section 3 outlines the baseline CMS event builder and how it can be implemented. The paper is summarised in Section 4.

---

[1] An average processing time of 300 ms was measured on a 2 GHz PentiumIII PC.

## 2. EVENT BUILDER STUDIES

Several architectures and switch technologies for the EVB have been evaluated by measurements with test benches and by simulation. The current EVB test-bench is based on 64 PCs acting as data sources and consumers and employing both Myrinet and Gigabit Ethernet technologies as the interconnect. In the case of Ethernet, protocols based on Layer-2 frames and on TCP/IP have been evaluated. Results from earlier work can be found in refs [5,6].

### 2.1. The EVB demonstrator

In the EVB test-bench, PCs are used to emulate Readout Units, Builder Units and an Event Manager. They implement the transmission of event data and control messages over the same physical builder network. RUs generate the event fragment data and BUs discard the event data once an event is fully assembled. The Level-1 Trigger is not emulated.

The EVM assigns a unique event identifier to label each event currently being processed by the event builder. The EVB protocol used is shown in Figure 2[2]. A BU sends event requests to the EVM, which replies with event identifiers of events allocated to this BU. In turn, the BU initiates the transfer of event data by requesting event fragments from each of the RUs. Once all fragments corresponding to one event have been collected by the BU, it sends a message to the EVM to clear the event identifier.

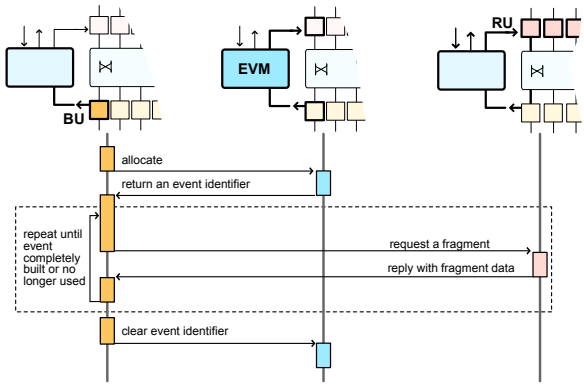

Figure 2: The EVB protocol.

The tests are throughput measurements for various configurations. As the trigger is not emulated, all measurements correspond to the saturation limit. The total volume of the EVB application payload transmitted is measured at each source and destination node. A run lasts in the order of minutes.

The EVB topology used is NxN, i.e. the number of sources and destinations are equal. Of particular interest is the scaling behaviour of the EVB. The performance of the NxN EVB scales with N, if the aggregate throughput grows linearly with N. Equivalently, the throughput per node is then independent of N. Another important measure is the load or utilisation efficiency of the network, i.e. the fraction used of the bisection bandwidth of the network.

The performance is studied with both fixed and variable size fragments. Variable sized event fragments are generated to mimic the sizes expected from CMS readout. The sizes are generated according to the log-normal distribution. The average and rms can be changed between runs.

### 2.2. Myrinet

Myrinet [7-8] is a high-speed, low-latency interconnect for clusters. A Myrinet network is composed of switching elements and network interface cards (NICs), connected by point-to-point bi-directional links. The effective link speed is currently 2 Gbit/s and the switching elements are based on 16-port crossbar chips. Myrinet employs wormhole routing and packets can be of arbitrary size. Flow control at the network link level guarantees the delivery of packets at the expense of probability of blocking in the switches. The NIC is composed of a control processor called "LANai" and 2 MB of SRAM. The LANai chip integrates a RISC processor core and DMA engines for the network link and host PCIbus. The SRAM serves as the network buffer and also as the code and data memory of the RISC. The RISC executes a customisable Myrinet Control Program (MCP), which supervises the DMA engines and implements a communication protocol.

A Myrinet network is essentially an input-queued switching fabric. Hence, depending on the traffic pattern, the throughput can be limited by head-of-line blocking. For random traffic, the utilization efficiency has an asymptotic theoretical limit of 59% for a single-stage network [9]. For multi-stage networks the utilisation is further reduced because of the increased probability of blocking at the additional stages. The wormhole routing implies that a blocked packet will also block other packets along its path back to the source, since there is no buffering in the switches. Event building can be more demanding than random traffic since it creates a high degree of congestion at the switch output nodes. Simulation studies of event building with variable-size fragments show that, depending on details of the fragment size distributions and the number of stages, a ~50% or lower utilisation is obtained. On the other hand, a close to 100% utilisation of the network can be reached, in principle, with barrel-shifter traffic shaping, described below.

---

[2] The full EVB protocol includes a message from the EVM to all RUs containing the association of event identifier and Level-1 Trigger information. As the Level-1 Trigger is not emulated in the EVB test bench, this part of the protocol is not implemented.

The EVB test-bench configuration is based on a 32x32 switch network. It comprises a half-populated Clos-128 switch connecting 64 PC nodes, acting as emulators of the RUs, BUs and EVM (see Figure 3). Details of the configuration can be found in Tables 3 and 4.

Table 3: PC hosts used in the EVB test-bench.

| Element | Equipment type |
|---|---|
| PC motherboard | SuperMicro 370DLE with ServerWorks LE chipset |
| CPU | Pentium-III, 750 MHz or 1 GHz |
| PCIbus | 64 bit / 66 MHz |
| Operating system | Linux 2.4 |

Table 4: Myrinet hardware used in the EVB test-bench.

| Element | Equipment type |
|---|---|
| NIC | M3S-PCI64B. LANai-9 based with 2 MB of local SRAM [8] |
| Switch | Clos-128 with M3-SW16-8S linecards [8] |

The software running on the PC nodes implements the event building protocol described above. A custom MCP and associated device driver for Linux has been developed implementing a zero-copy user-space message passing mechanism between hosts.

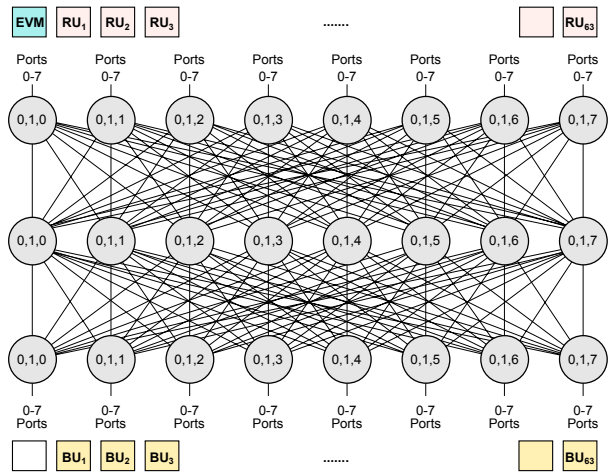

Figure 3: Myrinet Clos-128 network for a 64x64 EVB network. In the test bench only half of the ports are populated for a 31x31 EVB.

## 2.2.1. EVB without traffic shaping

The simplest scheme using Myrinet for the EVB is to apply no traffic shaping, i.e. to send packets on the network as soon as they are available. Depending on their size, the fragment data are spanned over one or more packets of Maximum Transfer Unit (MTU) size. No aggregation of event fragments into packets is performed, i.e. each event fragment starts in a new packet. The throughput per node for a 1x1, 8x8 and 32x32 EVB as a function of fragment size is shown in Figure 4. The 1x1 EVB reaches a maximum performance of ~225 MB/s[3], corresponding to ~90% utilisation for fragment sizes above ~2 kB. The sawtooth structure reflects the MTU size, which was set to 4 kB. As expected, the utilisation is reduced to ~50% for the 8x8 EVB using a single crossbar and further down to ~30% for the 32x32 EVB using two stages of crossbars.

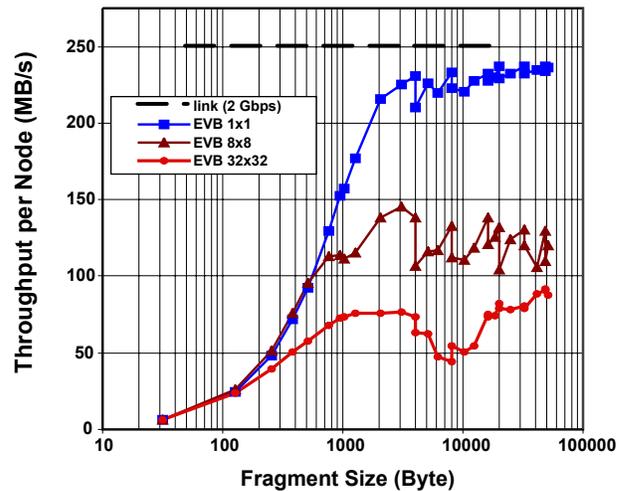

Figure 4: Throughput per node versus fragment size of the Myrinet EVB without traffic shaping.

## 2.2.2. EVB with Barrel Shifter

One can exploit the characteristics of event building traffic with the "barrel-shifter" technique [10], shown schematically in Figure 5. The basic idea is that the sequence of sends from each source to each destination follows the cyclic permutations of the destinations. Assuming fixed size event fragments and that event data are always available for sending to the destinations, the barrel-shifter uses 100% of the bandwidth of a non-blocking switch. A global synchronisation is required for the time-sliced operation of all sources. This is typically implemented using a central controller.

---
[3] 1 MB/s is defined as 10^6 B/s

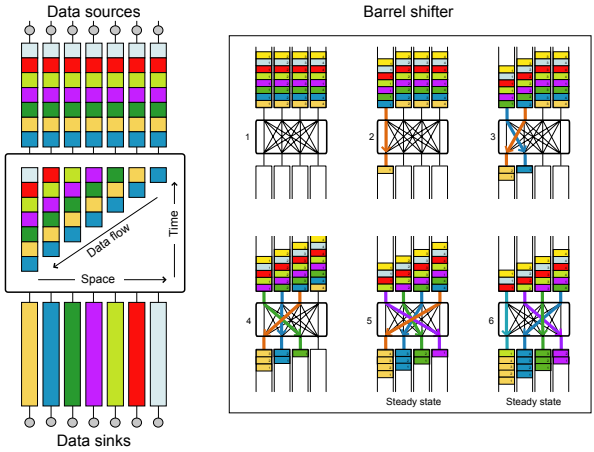

Figure 5: Schematic of the Barrel-Shifter traffic-shaping scheme. During the first time slot, only source 1 is sending data (to destination 1). During time slot 2, source 1 is sending to destination 2, while source 2 is sending to destination 1. During time slot *m*, source 1 is sending to destination *m*, source 2 to *m-1* and so on. After N time slots (where N is the number of sources) all sources are sending data to mutually exclusive destinations.

The barrel shifter has been implemented with Myrinet, taking advantage of the RISC on the NIC and the back-pressure flow control of the network. A custom MCP, running on the NIC, has been developed for this purpose. The basic idea is shown schematically in Figure 6. Each host on the source side has a separate queue for each possible destination. The MCP cycles through all destinations, copies the data by DMA to the NIC buffer and sends the data in carrier packets ("carriers") to all destinations in turn. The barrel shifter imposes fixed-sized carriers, but event fragments have variable sizes. In order to deal with this, fragment data buffers are spanned over or packed into the fixed sized carriers. The NIC of the sender spans and packs and the NIC of the receiver re-assembles.

In this implementation, no central controller is present for synchronisation. Instead, it takes advantage of the back-pressure flow control of Myrinet. After an initial synchronization step utilising the Myrinet network itself, all sources send through mutually exclusive paths to different destinations in a cycle. To keep this synchronisation, the sender NICs always emit packets, sending empty carriers when there is no data available for a particular destination.

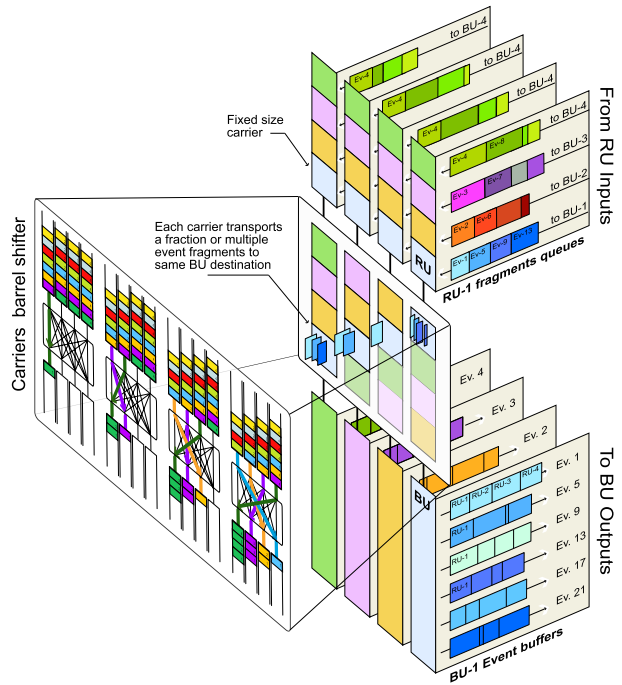

Figure 6: Schematic drawing of the Myrinet Barrel Shifter.

The barrel shifter technique works for single-stage and multi-stage Myrinet network fabrics and the throughput scales, in principle, to any size.

The barrel-shifter is implemented for the traffic in the top-bottom (RU/EVM to BU) direction. Note that no traffic-shaping is applied in the bottom-up (BU to RU/EVM) direction, used for small control messages. All measurements were done with the barrel-shifter carrier size set to 4 kB, corresponding to a transfer time of about 17 µs.

A configuration with a 32x32 network implementing a 31x31 EVB has been tested. Note that 1 out of the 32 barrel-shifter ports is connected to the EVM (see Figure 3). The throughput per node as a function of fragment size is shown in Figure 7. In the case of fixed size fragments, it reaches a plateau of about 230 MB/s for sizes above 6 kB. If needed, the plateau can be reached for lower fragment sizes by aggregating the control messages. The performance for variable size event fragments, is also shown in Figure 7, and it can be seen that the throughput per node reduces with increasing rms, as expected. For the nominal case[4] of variable size fragments with an average size of 16 kB and a rms of 8 kB, a throughput per node of about 215 MB/s is achieved.

---

[4] The nominal case of 16 kB fragments corresponds to the parameters of the baseline EVB design (see section 3.1).

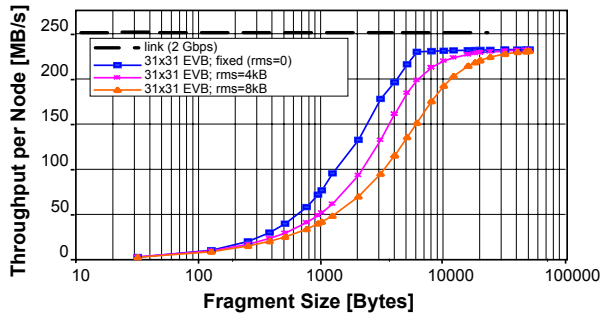

Figure 7: Throughput per node versus average fragment size of the Myrinet 31x31 EVB with barrel shifter. The series with variable sizes corresponds to log normal distributions.

In order to investigate the scaling of the performance, network configurations of 8x8, 16x16, 24x24 and 32x32 have been tested. As one port of the barrel shifter is connected to the EVM, they correspond to a 7x7, 15x15, 23x23 and 31x31 EVB, respectively. The results of these measurements are presented in Figure 8. The throughput per node for an NxN EVB varies as N/N+1, because N out of N+1 cycles of the barrel shifter are used to transfer event data.

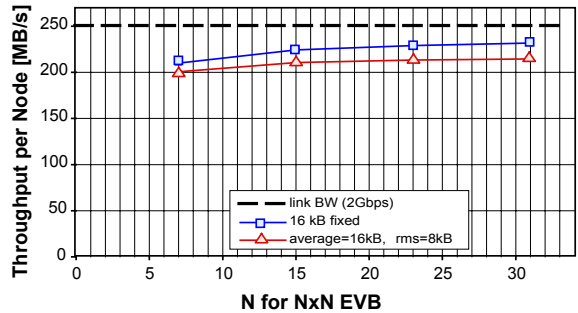

Figure 8: Throughput per node versus NxN for Myrinet EVB with barrel shifter.

## 2.3.  Gigabit Ethernet

The term Ethernet refers to the family of LAN products covered by the IEEE 802.3 standard protocol. Four data rates are currently defined for operation over twisted-pair cables or optical fiber. Gigabit Ethernet with a bandwidth of 1 Gbit/s in each direction is used for the EVB test bench.

An Ethernet network is composed of switching elements and NICs connected by point-to-point bidirectional links. Store-and-forward switches permit the interconnection of links with different data rates. Frame size is limited to 1500 B of payload data. The NIC is in general not programmable; but the switches are configurable and support extensions to the standard such as VLAN, which effectively divides a switch into several smaller switches.

The switches are in general non-blocking, i.e. the switch can support the aggregate rate of all the ports. However, packets can still get lost because of capacity limitations in the store and forward memories. Flow control is optional and is implemented through the use of pause frames generated by the Media Access Control (MAC) layer of the receiving NIC. Flow control is generally implemented between NIC and switch port, but not always through the switch back to the source NIC.

An important design choice for the EVB concerns the communication protocol. In particular whether to use Ethernet Layer-2 frames or a reliable high level protocol, such as TCP/IP. Both options have been studied.

A 31x31 EVB based on Gigabit Ethernet has been set up. The nodes are 63 PCs emulating the RUs, BUs and EVM. Details on the configuration can be found in Tables 3 and 5. A fully populated FastIron-8000 switch comprises 8 modules with 8 Gigabit Ethernet ports each, connected to a crosspoint backplane. Each module contains 2 MB of shared memory to buffer packets. The bandwidth of the memory system and backplane is dimensioned such that the switch is fully non-blocking. For our tests, the switch is configured such that each port can buffer up to 63 packets. The switch model used does not support Jumbo frames. Therefore, all present studies are restricted to standard frame sizes (MTU=1500 B).

Table 5: Ethernet hardware used in the EVB test-bench.

| Element | Equipment type |
|---|---|
| NIC | AceNIC from Alteon [11] running standard firmware |
| Switch | FastIron-8000 from Foundry Networks [12]. |

### 2.3.1.EVB with Layer-2 Frames

Experience so far has shown that, typically, large switches do not implement endpoint-to-endpoint flow control through the switch. Hence, packets can get lost due to congestion. With destination based traffic shaping the congestion can be reduced. In this scheme, one relies on the destinations to send their requests for data to the sources in an order that inherently avoids clashes. The EVB is driven by the destinations, as the BUs send requests for data only once they have buffers available. The simplest way to shape the traffic at the destination is to assume that the destination sends a request to N RUs sequentially: a request to a RU (e.g. "send event 45") is then followed by the actual data transfer of the data from event 45, at which point the request for the same event is sent to another RU, and so on, as shown in Figure 9. This scheme avoids congestion at the switch outputs if a BU builds a single event at a time. However, no data is transmitted into the BU link during the time the request message is sent and processed. This results in a reduced utilisation of the switch network, because of  network latency and processing times. In order to use the links efficiently, several events have to be built concurrently (set to 32 events per BU in the test). Packet loss can then occur occasionally if several RUs send instantaneously to the same destination port, which will drop a packet if its buffer capacity of 63 packets is exceeded. The protocol has been augmented to recover from this occasional packet loss. It is based on sequence numbers, time-outs and a retry mechanism.

The EVB application software running on the PCs implements the event building protocol discussed above, using Layer-2 frames to transmit data. No aggregation of event fragments into packets is performed. A library has been developed to transmit Layer-2 frames from user space with zero-copy. The Linux driver for the NIC was modified for this purpose.

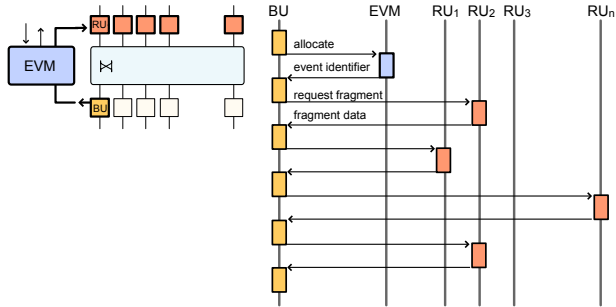

Figure 9: Protocol for the Ethernet EVB with Layer-2 Frames.

The throughput per node as a function of the fragment size is shown in Figure 10. Also shown is the calculated maximum performance based on the link bandwidth and header sizes. The saw-tooth structure reflects the Ethernet MTU size. For fragment sizes up to roughly 8 kB, the throughput is reduced due to the packet handling overhead for control and data messages[5]. For medium fragment sizes in the range 8 kB to 20 kB, close to maximum performance is achieved. For fragment sizes larger than 20 kB, significant packet loss leads to inefficiency due to retransmission.

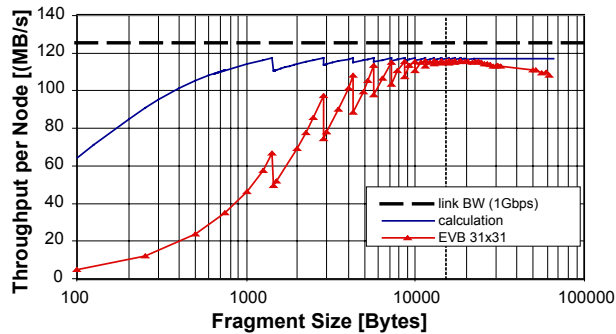

Figure 10: Throughput per node versus fragment size for the 31x31 one-rail Gigabit Ethernet EVB.

---

[5] A 15x15 EVB setup based on different hardware and software with lower overhead is described in [6].

The performance for variable sizes has been studied as well. As expected, only a small degradation with respect to the case of fixed size has been observed (see Figure 11).

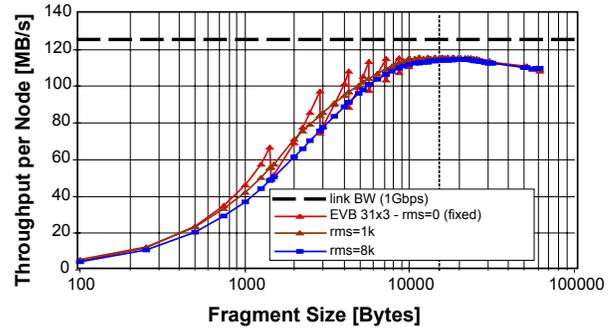

Figure 11: Throughput per node versus average fragment size for the 31x31 one-rail Gigabit Ethernet EVB. The fragment sizes are generated according to log-normal distributions.

The size of the EVB is varied from 3x3 to 31x31 in Figure 12 for the nominal case of variable size fragments with an average of 16 kB and an rms of 8 kB. It can be seen that the throughput per node scales with the size of the configuration and a value of 115 MB/s is reached.

A two-rail 15x15 EVB has been assembled with the same hardware by equipping all nodes with two NICs. Point-to-point measurements over two links show that the transmission rate is almost a factor two higher than with a single link, indicating that the transfers via the two NICs largely overlap. The resulting EVB performance for nominal fragment sizes is compared to the one-rail configuration in Figure 12. The throughput per node scales with the size of the configuration and a value of 220 MB/s is achieved.

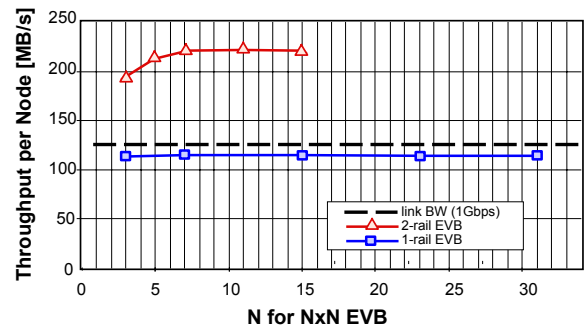

Figure 12: Throughput per node versus NxN for one-rail and two-rail Gigabit Ethernet EVB. The fragment sizes are generated according to a log-normal distribution with an average of 16 kB and an rms of 8 kB.

### 2.3.2. EVB with TCP/IP

Using TCP/IP rather than Layer-2 frames for event building has a number of advantages. Foremost, it provides a reliable transport service that removes the need for the event building application to detect and deal with lost packets. TCP/IP also provides flow control and congestion control. Flow control tries to avoid buffer overruns at end points, whereas congestion control tries to prevent packet loss in intermediate nodes (switches or routers). Besides these technical merits, TCP/IP also provides a solution based on standard software. However TCP/IP is more demanding on host resources[6] and the scaling properties for event builder traffic at high load are still unknown.

The TCP/IP option has been evaluated using the same hardware test-bench. TCP/IP performance is highly dependent on NIC and host CPU performance and on the TCP/IP protocol stack implementation of the host-OS. A one-way streaming test between two PCs of the EVB test bench yields a maximum throughput of about 88 MB/s for message sizes above 800 B, corresponding to 70% of the link bandwidth.

The EVB application software running on the PCs implements the event building protocol discussed above, using TCP socket connections to transmit data. Each instance of the RU, BU or EVM application runs as a single process on its host. Synchronous I/O multiplexing is implemented using the select mechanism to send and receive data via socket connections to all remote applications. The control messages are grouped for efficiency reasons. The request messages from BU to RU and the allocation or clear messages between BU and EVM aggregate 8 and 16 logical messages, respectively. Furthermore, each BU builds 64 events concurrently.

Studies were done with the 31x31 EVB configuration. The throughput per node as a function of the fragment size is shown in Figure 13. For a size of 16 kB the achieved throughput per node is about 75 MB/s. This throughput corresponds to 85% of the measured one-way streaming throughput (88 MB/s). It should be noted that due to this limitation of the host and/or NIC, the switch utilisation is only about 60%.

Measurements of the performance for variable sizes shows no considerable degradation with respect to the case of fixed size.

The size of the EVB is varied from 3x3 to 31x31 in Figure 14 for the nominal case of variable size fragments with an average of 16 kB and an rms of 8 kB. It can be seen that the throughput per node scales approximately with the size of the configuration and a value of about 75 MB/s is achieved.

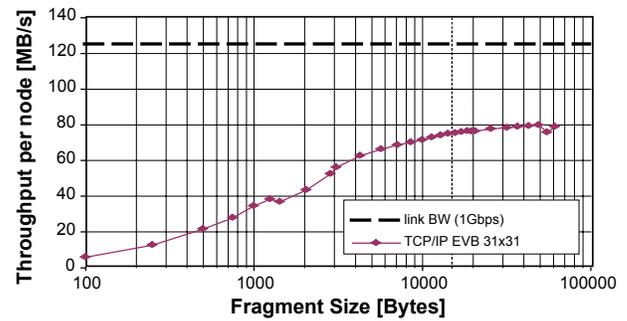

Figure 13: Throughput per node versus fragment size for a TCP/IP 31x31 EVB.

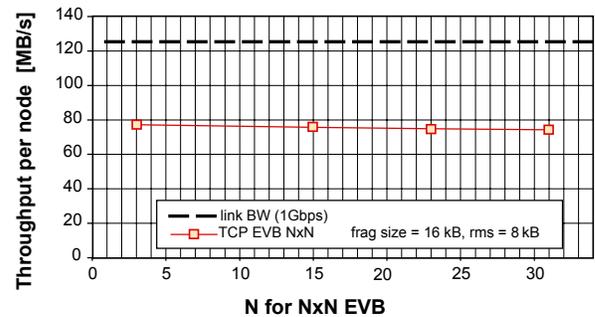

Figure 14: Throughput per node vs. NxN for a TCP/IP EVB. The fragment sizes are generated according to a log-normal distribution with an average of 16 kB and an rms of 8 kB.

A higher throughput can be obtained with more recent hardware. For example, preliminary measurements on PCs with 2 GHz Pentium-IV Xeon processors and appropriate NICs yield a value of about 100 MB/s per node for a 10x10 EVB configuration.

## 2.4. Comparison

The results obtained with the 32x32 EVB test setup for the Myrinet, Gigabit Ethernet Layer-2 and TCP/IP technologies are summarised in Table 6. The achieved utilisation of the network is roughly 90%, 90% and 60%, respectively, for fragment sizes above 10 kB. This implies for the CMS event builder with a required 100 GB/s aggregate throughput a switch network consisting of one rail, two rails and three rails of a 512x512 port switch, respectively. The extrapolation assumes scaling from the 32x32 test bench to a 512x512 configuration.

Table 6. Summary of the EVB test bench results and extrapolation to the CMS 100 GB/s EVB.

|            | Myrinet-2000 | GbEthernet Layer-2 | GbEthernet TCP/IP |
|------------|--------------|--------------------|-------------------|
| Test bench | 32x32        | 32x32              | 32x32             |
| Port speed | 2 Gbit/s     | 1 Gbit/s           | 1 Gbit/s          |
| Utilisation | ~90% (BS)   | ~90%               | ~60%              |
| CPU load   | Low          | Medium             | High              |
| 100 GB/s EVB | 512x512    | 2*512x512          | 3*512x512         |

---

[6] According to an empirical rule, 1 Hz of CPU is required per bit/s transmitted.

## 3.  THE CMS TWO-STAGE EVB DESIGN

Assuming an Event Builder with ~500 data sources and ~500 data consumers, appropriately interconnected by a large switch network, a design choice whether to perform the event building in one or more steps has to be considered. Related to this is the number of stages in the switch network.

• The first possibility is a system which contains a single switching stage that connects 500 data sources to 500 data consumers via a single fabric. The fabric may consist of multiple stages of switches, but, no explicit buffering is added in between the intermediate stages beyond that already provided by the switches themselves. The event building is done in a single step.

• The second possibility is a system in which there are two switching stages which are separated by a buffering layer. This intermediate layer performs a first level of event assembly.

The advantage of such an intermediate stage is the resulting increase in the data size along with the decrease in rate of the event fragments that have to be transported through the switch. Furthermore, whereas the single fabric configuration demands scaling of the network to a very large number of ports (500x500), the two-stage configuration employs smaller switching fabrics, which makes it easier to achieve a high utilisation of the network.

The second configuration has been chosen as the baseline for the CMS event builder. A major issue is the modularity of the system and its effect on scalability. The modularity of the system with an intermediate buffering stage is higher and allows for a phased installation of the DAQ. In addition, this option decouples the choice of technology for the switches used for the two stages.

### 3.1.  EVB conceptual design

The conceptual design of the baseline Event Builder is shown in Figure 15 and consists of a FED-builder stage and a RU-builder stage.

The Front-End Drivers are connected to Front-End Readout Link (FRL) modules, which can merge up to two FED inputs. This merging capability is useful for the detector channels whose data volume is well below the nominal average. As a result, the ~700 FEDs are concentrated into ~500 switch inputs, each delivering event fragments of about 2 kB on average.

The FRLs in turn, are connected to Readout Units via a small (8x8) switch network - labeled as "FED Builder" in Figure 15. For any event in the system, there are eight potential destinations, as far as the front-ends are concerned. On the other hand, each one of the RUs is a point of merging of eight fragments into a super-fragment. Given the parameters of the CMS system, each RU will thus build super fragments with a mean size of 8x 2 kB=16 kB of data. There are a total of 64 FED Builders, one per RU. The destination assignment in the FED Builder stage is based on the trigger number stored in the FED data record. The algorithm to select the destination is in the simplest case, a round-robin across all outputs, implemented as modulo 8 of the trigger number.

Once an event has been forwarded from the FRLs to the 64 RUs which are connected to the second stage RU-builder network, the Readout Units proceed to send their super-fragments for this event to a BU via the large (64x64) data network. The BUs receive the 64 super fragments and build the final single-event. The destination assignment in the second stage is done dynamically by an Event Manager, providing load balancing between BUs and associated FUs. The collection of 64 RUs and BUs, the EVM and the switch network connecting them is referred to as a "RU Builder". There are eight RU Builders in the full CMS system.

As a result of the independence between the two stages, the second stage may consist of one to eight RU Builders. The smallest configuration, having a single RU Builder, is capable of handling a 12.5 kHz trigger rate. In this case, all FED Builders only use one out of the eight output ports of the FED Builder switch. A configuration with two RU Builders uses two of the output ports. A possible destination assignment is in this case to send events with odd trigger numbers to RU Builder number 1 and events with even trigger numbers to RU Builder number 2. This multiplexing of the FED Builders to different RU Builders establishes a natural split of the Event Builder into eight independent DAQ systems. The system's scalability in multiples of the basic unit, the RU Builder, is built into the design. It is foreseen to install four RU Builders at the start of data taking in 2007, capable of handling a 50 kHz trigger rate.

As a consequence of the two-stage EVB design with 64 8x8 switches in the FED Builder stage and 8 64x64 switches in the RU Builder stage, the RU and BU nodes need to operate at a rate of only 12.5 kHz with average super fragment sizes of 16 kB, rather than at 100 kHz with average fragment sizes of 2 kB. This reduces the rate of fast control messages to control the event building process.

Furthermore, the system has an increased level of redundancy. An entire RU Builder may cease to operate during data-taking, yet the system can continue to run, albeit with lower (i.e. 7/8 of the full) performance. In addition, a RU Builder may be dedicated to testing in parallel with the normal operation of the system.

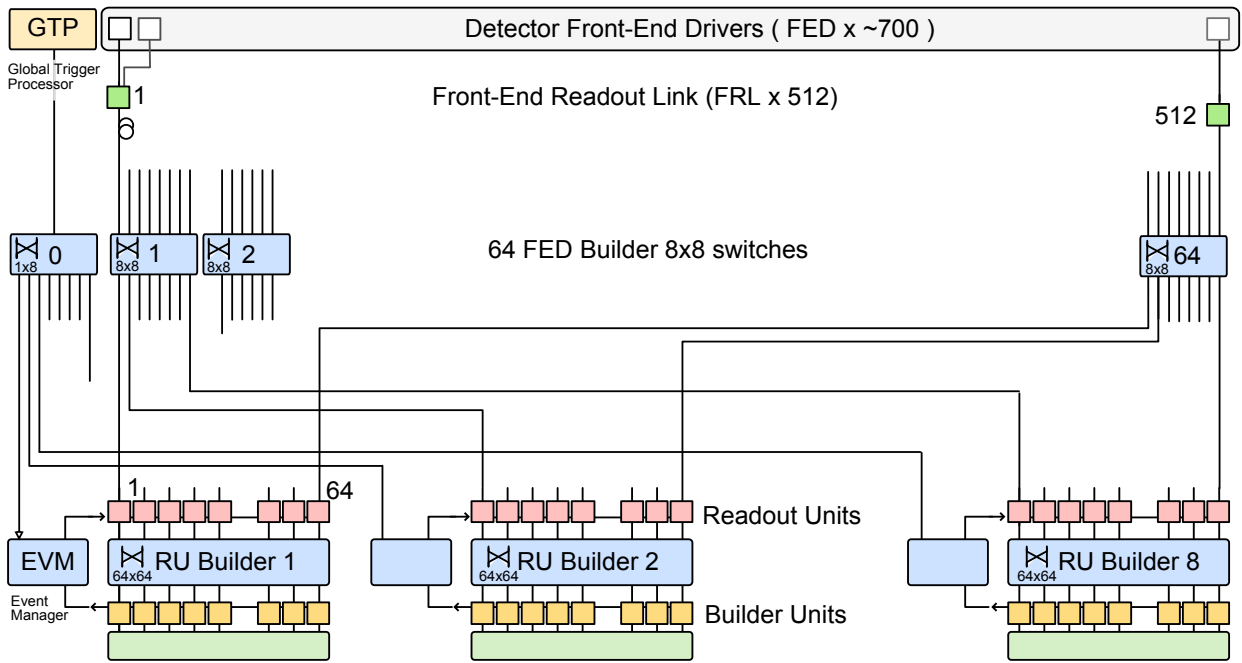

Figure 15: Conceptual design of the CMS Event Builder. The FED Builder stage consists of 64 FED Builders (only number 1,2 and 64 are shown) and one special FED Builder interfacing the Global Trigger Processor to the Event Managers (number 0). The RU Builder stage consists of 8 RU Builders (only number 1,2 and 8 are shown). The connection between FED Builder switches and Readout Units is as follows: Output port $i$ ($i$=0..8) of FED Builder switch $j$ ($j$=1..64) is connected to RU number $j$ in RU builder number $i$.

## 3.2.   Implementation

Possible implementations for the components of the EVB are outlined below.

### 3.2.1. Front-End Readout Link

The FRL module is a custom FPGA-based design with up to two SLINK-64 [13] input ports and a PCI(-X) output port. The PCI(-X) output port hosts a NIC and ensures flexibility for the technology used for the FED Builder network. A prototype FRL is currently under development.

### 3.2.2. Readout Unit and Builder Unit

Both RU and BU nodes are required to sustain an I/O throughput of at least 200 MB/s for both input and output and to buffer event data for several seconds, corresponding to ~1 GB of memory. This requirement can be met with today's server PCs, which feature at least two independent PCI(-X) buses and a main memory system with a capacity of O(GB) and a bandwidth of O(GB/s). Furthermore, the available computing power on these multi-CPU computers allows the implementation of the EVB logic in high-level software and possibly the use of high level networking protocols.

The availability of cost-effective PCs ensures flexibility for changes in EVB protocols, hardware setup and networking equipment. The RU and BU nodes need to operate at a rate of 12.5 kHz with average super fragment sizes of 16 kB. RUs and BUs implemented in software using the XDAQ framework [3,14] have been shown to meet the performance requirements.

### 3.2.3. FED builder network

The 8x8 FED Builder is required to handle a 100 kHz rate of fragments with sizes ranging from a few hundred bytes to a few kilobytes and combine them into super-fragments.

The implementation of the FED Builder with Myrinet has a number of attractive features. Myrinet supports reliable transport with back pressure at the hardware level, the NICs are programmable and have sufficient memory to smooth bursty arrival of data from the FRL. The network utilisation will be roughly 50% when transporting variable size fragments. Given the link bandwidth of 2 Gbit/s, the switch capacity has to be doubled to sustain an average of 200 MB/s per port. LANai-10 based NICs with two links per NIC can be used to construct a two-rail network, where each link of the NIC is connected to two independent 8x8 crossbar switches (see Figure 16).

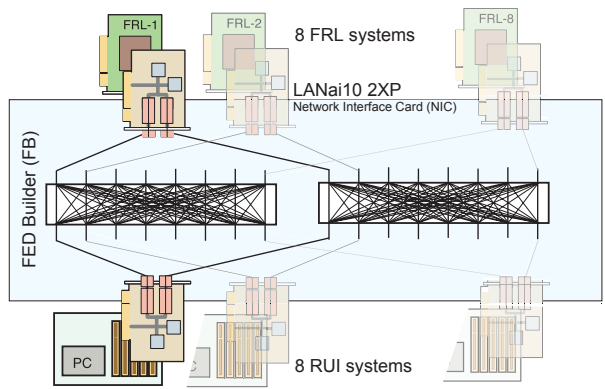

Figure 16: FED Builder based on Myrinet.

The FRLs with their NICs are located in the cavern close to the FED crates, whereas the switches are located at the surface in the DAQ building. The Myrinet links from the FRL NICs to the switches are standard optical multi-mode fibers. The required distance is within the maximum fiber length of 200 m.

The results from the FED Builder prototype combined with simulation studies show that the FED Builder can be implemented with this configuration and can be designed to deliver super-fragments in sequential order to the RU at the required 12.5 kHz rate with a low overhead on the RU.

### 3.2.4. RU builder network

The 64x64 RU Builder is required to handle a 12.5 kHz rate of 16 kB fragments. This corresponds to an aggregate effective throughput of 13 GB/s. The test benches described in section 2 already correspond to half the required size and together with simulation results [3] show that the requirements can be met with Myrinet and most likely also with Ethernet technology.

Use of Myrinet for the RU Builder with a large number of ports necessitates traffic shaping due to the multiple stage crossbar fabric involved. The presented measurements on the half-sized RU Builder (32x32) prove that Myrinet can perform event building in barrel shifter mode at near 100% utilisation of the network. Simulation estimates show that a one-rail Myrinet configuration will meet the requirements of a full RU Builder.

The RU Builder can also be implemented with Ethernet technology. The required 200 MB/s sustained throughput per node can be reached with a two-rail Gigabit Ethernet configuration, assuming an effective throughput above 80% of the wire speed. In this configuration, each node accommodates two ports, with each port connected to one of the two rails. The network for each rail is provided by a 128-port switch. As an alternative to a single chassis 128-port switch, the switch network can possibly, be provided by a multi-stage arrangement of switches with a smaller number of ports per chassis.

The demonstrator has shown that it is feasible with a lightweight protocol using Layer-2 frames to operate an event builder near 100% utilisation of the network. The performance is critically dependent on the amount of internal buffer memory of the switch to ensure a low packet loss probability. This solution requires application level traffic shaping and error recovery.

The TCP/IP based solution would leverage the full advantage of standard Ethernet hardware and software with a reliable transport layer. The equipment available for prototyping was "state-of-the-art" in 2001. It could operate TCP/IP at 60% of the available bandwidth in event building with a (32x32) one-rail and a half-sized RU Builder. Further investigations with equipment of higher performance and larger configurations are needed to evaluate the feasibility of the TCP/IP option.

### 3.3.  Performance

The performance of the full EVB with the combined FED Builder and RU Builder stages has been estimated by simulation [3]. Both stages have been studied separately considering the missing element as a pure sink or source, respectively. This assumption that both stages are decoupled is justified as the memory in the RU can absorb the fluctuations of the output of the FED Builder. The overall EVB performance is clearly determined by the minimum of the performance of the FED Builder and RU Builder stages. A prediction is made for the maximum trigger rate as a function of the number of RU Builders, assuming balanced and uncorrelated inputs.

Implementations with Myrinet technology have been studied in detail. The simulation has been validated by comparison with measurements on LANai-9 based prototype setups. The expected performance for LANai-10 based hardware is shown in Figure 17. The FED Builder is based on the two-rail configuration described in Sect. 3.2.3, whereas the RU Builder is based on a barrel shifter with a one-rail network. The capacity of the set of RU Builders increases linearly with the number of RU Builders (by construction) and will hold irrespective of the technology used for the RU Builder. The FED Builder performance does not scale linearly because of blocking in the crossbar. However, the FED Builder stage always exceeds the capacity of the RU Builder stage. A maximum trigger rate of 110 kHz can be reached with the full EVB system.

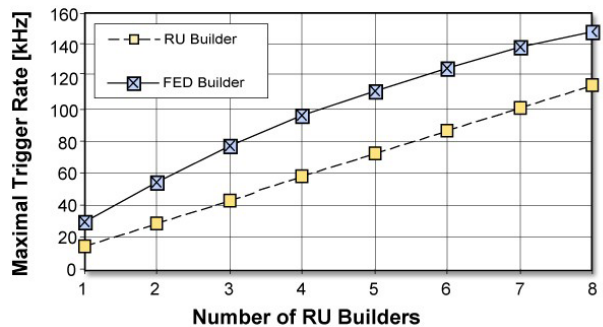

Figure 17: The maximum trigger rate as a function of the number of RU Builders in the DAQ system for a two-rail FED Builder and a one-rail Myrinet Barrel Shifter RU Builder. Inputs are assumed to be balanced and uncorrelated. Inputs are generated according to a log-normal distribution with an average of 2 kB (16 kB) and an rms of 2 kB (16 kB / √8), for the FED Builder (RU Builder), respectively.

## 4. CONCLUSION

The presented DAQ design fulfills the following major CMS requirements:

- Handling a 100 kHz trigger rate
- An event builder assembling full events and providing a scalable structure that can go up to 100 GB/s.
- A High Level Trigger by commodity processors in a single farm having access to the full event data.

It has been demonstrated that the EVB can be implemented with technology available in 2003. In the case of the FED Builder the requirements can be met with Myrinet and in the case of the RU Builder they can be met with both Myrinet and Gigabit Ethernet. The EVB architecture is flexible enough to be implemented with other network technologies as well.